\documentclass[12pt]{article}
\usepackage{graphicx}
\usepackage{epsfig}
\usepackage{amsbsy}

\begin{document}
\baselineskip 10mm

\centerline{\large \bf Thermal Desorption of Hydrogen From Graphane }

\vskip 6mm

\centerline{L. A. Openov$^{*}$ and A. I. Podlivaev}

\vskip 4mm

\centerline{\it Moscow Engineering Physics Institute (State
University), 115409 Moscow, Russia}

\vskip 2mm

$^{*}$ E-mail: LAOpenov@mephi.ru

\vskip 8mm

\centerline{\bf ABSTRACT}

The process of hydrogen desorption from graphane (graphene sheet saturated by hydrogen
adsorbed from both sides) has been studied using the method of molecular dynamics. The temperature
dependences of the time of desorption onset for various hydrogen coverages on graphene are
calculated and the corresponding activation energies in the Arrhenius equation are determined.
It is established that graphane exhibits a rather high thermal stability that makes possible its use
in two-dimensional electronics even at room temperature. For the same reason, graphane can hardly be
considered as a promising hydrogen storage material for fuel cells. 

\newpage

Graphene, representing a monolayer sheet of carbon atoms, draws the attention of both basic science
(as a medium featuring massless Dirac fermions) and practical applications (as a material for
nanoelectronics) [1]. It is also of interest to study the properties of graphene derivatives
(e.g., in the form of nanoribbons [2]), including those obtained by its chemical modification with
polymers, nitrogen, hydrogen, etc. Sofo et al. [3] theoretically predicted the existence of
graphane, representing a graphene sheet saturated by hydrogen adsorbed from both sides, and quite
recently this prediction was experimentally confirmed [4].

Graphane (unlike graphene) is a dielectric and, in principle, it can also be used in nanoelectronics.
The relatively large content (7.7 mass \%) of hydrogen in graphane suggests that it probably can be
used in hydrogen power engineering. In assessing the possibilities of the practical usage of
graphane, it is necessary first to study its thermal stability with respect to hydrogen desorption
and thus determine the range of permissible temperatures for potential applications. This Letter
presents the results of an investigation of the desorption of hydrogen from graphane, which was
numerically simulated using the method of molecular dynamics.  

The initial sample of graphene was modeled by a hexagonally packed monolayer fragment consisting of
54 carbon atoms. In order to reduce the influence of the finite size of the sample (edge effects),
the dangling bonds of $sp$-hybridized edge carbon atoms were saturated by hydrogen. This fragment
was converted into graphane by attaching one hydrogen atoms to each carbon atom alternatively from
the two sides of the monolayer. As a result, we obtained a C$_{54}$H$_{72}$ model cluster of
graphane schematically depicted in Fig. 1. In this cluster, all atoms occur in the $sp^3$-hybridized
state [4] (in contrast to the  $sp^2$-hybridization in graphene) and are shifted in the transverse
direction (off-plane) by 0.2 - 0.3 {\AA} depending on the position relative to the center of the
cluster. The C-H and H-H bond lengths in the graphane fragment are
1.10 and 1.50 - 1.53 {\AA}, respectively. Note that it is necessary to distinguish between 18
hydrogen atoms used to passivate the sample edge and 54 hydrogen atoms forming the graphane
structure: the former H atoms are present in both graphene and graphane models, whereas the latter
are present only in graphane; in macroscopic graphane samples of increasing size, the relative
fraction of the former H atoms tends to zero, while that of the latter tends to unity. 

The aim of this study was to determine the temperature dependence of the time necessary for the
detachment of one H atom (of the total of 54 hydrogen atoms entering into the graphane structure)
from the C$_{54}$H$_{72}$ cluster. This corresponds to the removal of a 1/54  $\approx 2$ \%
fraction of H atoms from a macroscopic sample of graphane, that is, to the formation of a relatively
large number of defects (sufficient to significantly change the electrical characteristics). This
task was solved using the method of molecular dynamics [5, 6]. At the initial moment, all atoms were
randomly imparted velocities and displacements so that the total momentum and moment of the entire
cluster were zero. Then, the forces acting on atoms were calculated and the classical Newton
equations of motion were numerically integrated using the velocity Verlet method with a temporal step
of $t_{0}=2.72\times10^{-16}$ s. The total system energy (a sum of the potential and kinetic
energies) remained unchanged, which corresponded to a microcanonical ensemble (a system not in
thermal equilibrium with the environment) [5, 6]. The relative motion of atoms was characterized by
the dynamic temperature $T$ that was determined from the following relation:
$\langle E_{kin} \rangle=\frac{1}{2}k_{B}T(3n-6)$ where $\langle E_{kin} \rangle$ is the kinetic
energy averaged over the time, $k_B$ is the Boltzmann constant, and $n$ is the number of atoms in the
system. This approach was successfully used previously for the qualitative explanation of
experimental data on the fragmentation of C$_{60}$ fullerene [6]. 

An important question is how to calculate the forces acting on atoms at each step of molecular
dynamics. In fact, this procedure reduces to calculating the total system energy as a function of
the coordinates of component atoms. Solving this task by the ab initio methods requires considerable
computational facilities and allows the evolution of a system involving $\sim 100$ atoms to be traced
for a rather short time on the order of $\sim 10$ ps, which is insufficient to accumulate a
necessary statistics. For this reason, we employed a nonorthogonal tight-binding model [7], which
reasonably compromise between the strict ab initio methods and oversimplified approaches based on
classical interatomic interaction potentials. The proposed model rather well describes both small
carbon clusters (e.g. fullerenes [7]) and hydrocarbon structures (e.g., cubane C$_8$H$_8$ [8]) and
macroscopic systems [7]. We believe that the method of molecular dynamics in combination with a
tight-binding potential provides a quite adequate description of the thermal stability of
graphane, similar to the results obtained by applying the same approach to cubane [8].

Evolution of the C$_{54}$H$_{72}$ cluster was traced to the moment of detachment of the first H
atom at various temperatures in the range of $T = 1300 - 3000$ K. The results presented in Fig. 2
show that as the temperature decreases, the desorption onset time $\tau$ exponentially increases
from $\sim 0.1$ ps to $\sim 10$ ns. The plot of ln($\tau$) versus 1/$T$ is well approximated by a
straight line, which indicates that the process of hydrogen desorption can be described in terms of
the following standard Arrhenius relationship:
\begin{equation}\label{1}
\tau^{-1}(T)=A\cdot\exp\left[-\frac{E_{a}}{k_{B}T}\right]
\end{equation}
where $A$ is the temperature-independent (or weakly dependent) frequency factor and $E_a$ is the
activation energy (which can be determined from the slope of the straight line in Fig. 2).
Statistical analysis of the results of modeling yields $E_a=(2.46 \pm 0.17)$ eV and
$A=(2.1 \pm 0.5)\times 10^{17}$ s$^{-1}$. By substituting these $E_a$ and $A$ values into formula
(1), we can evaluate $\tau$ at low temperatures, which are inaccessible for computer simulations
because of extremely long necessary computation times. At $T = 300$ K, this estimation yields a
macroscopic value of $\tau \sim 10^{24}$ s, which is evidence for the possibility of using
graphane in nanoelectronic devices operating at room temperature. As the temperature increases to
600 K, the value of $\tau$ drops to $\sim 1000$ s. These results do not contradict the recent data [4], according
to which the complete desorption of hydrogen from graphane was achieved by annealing it for one day
in argon at $T=700$ K. It should be noted that, in view of the exponential character of relation (1),
any refinement of the $E_a$ value can lead to rather significant changes in the values of $\tau$.
Therefore, the above values of $\tau$ at different temperatures should be considered as rough
estimates.
 
A detailed analysis of evolution of the model C$_{54}$H$_{72}$ cluster showed that the desorption of
hydrogen from various parts of the cluster takes place at approximately equal probability . This fact
agrees with the results of calculations of the heights $U$ of energy barriers for the desorption of
H atoms. These values were determined by studying the potential energy hypersurface of the given
cluster as a function of the coordinates of component atoms (see [9]). The barrier height $U$
amounts to 2.4 eV at the center of the cluster and increases to 2.7 eV at the periphery. As expected,
these values are close to the desorption activation energies $E_a$. It should be noted that, in the
course of simulations, we observed two cases of the desorption of H$_2$ molecules rather than atomic
hydrogen. 

In order to study the dependence of the rate of desorption on the degree of graphene coverage by
hydrogen (graphane corresponds to the total coverage, whereby the number N$_H$ of nonpassivating
hydrogen atoms is equal to the number N$_C$ of carbon atoms in the sample), we have also determined
$\tau$ as a function of $T$ for two other systems representing graphene with a half coverage
(N$_H$ = N$_C$/2, C$_{54}$H$_{45}$ cluster) and with a single adsorbed H atom (N$_H$ = 1,
C$_{54}$H$_{19}$ cluster). The corresponding values of the desorption activation energy
($E_a=(1.74 \pm 0.17)$ and $E_a=(1.86 \pm 0.14)$ eV, respectively) proved to be smaller compared to
the value obtained for graphane, in agreement with the lower barriers for desorption
($U=(1.2-1.9)$ eV, increasing toward the periphery) and with the general tendency of a decrease in
the thermodynamic stability of hydrocarbons with decreasing relative content of hydrogen [3]. These
results indicate that the time of the total desorption of hydrogen from graphane is determined by
the initial stage, in which the hydrogen content decreases by (1 - 10) \%.

It is interesting to note that, in simulating the C$_{54}$H$_{45}$ and C$_{54}$H$_{19}$ clusters, we
have repeatedly observed the fracture of the carbon framework much before the hydrogen desorption
event. This behavior agrees with the data of experiments [4], in which an increase in the
temperature of graphane annealing above a certain level led to damage of the graphene framework.
Taking into account that (i) this effect is not observed in the C$_{54}$H$_{72}$ cluster and (ii)
the rate of hydrogen desorption increases with decreasing content of H atoms, it may be recommended
to perform the initial stage of annealing at a higher temperature and then slightly cool the sample.

In conclusion, let us briefly consider the possibility of using graphane in fuel cells for electric
car engines. In addition to meeting the requirements of high hydrogen content (more than 6 mass \%)
and stability at room temperature, it is necessary that the fuel-cell material would be capable of
rapidly (within $\sim 1$ s) and almost completely release hydrogen at a temperature not exceeding
400 K [10]. According to the results of our investigation, graphane does not obey the last
requirement. This is probably related to the strong covalent C-H bonds which, on the one hand,
ensure a high thermal stability of hydrogen chemisorbed on carbon nanostructures [10] and, on the
other hand, sharply decrease the rate of hydrogen desorption. Therefore, the most promising
directions of the practical use of graphane are apparently related to nanoelectronics. In this
context, it will be of interest to study the phenomenon of thermoactivated migration of hydrogen
through the graphene/graphane interface. Should this boundary remain atomically sharp at not very
low temperatures, it will be possible to attempt at creating various nanoelectronic devices by
selectively adsorbing hydrogen on graphene sheets and nanoribbons.

\newpage
\vskip 20mm
\includegraphics[width=\hsize,height=15cm]{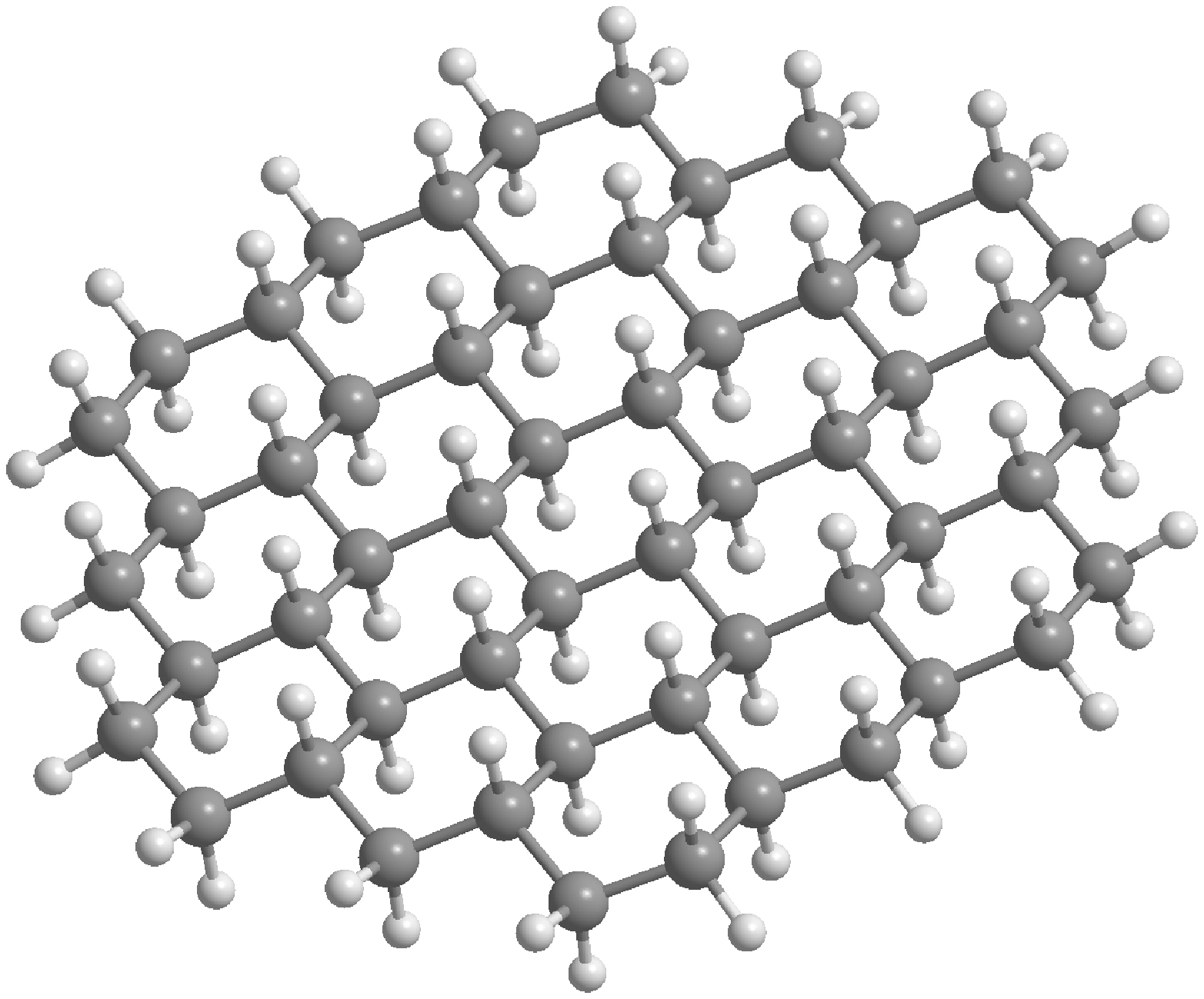}
\vskip 40mm
Fig. 1. C$_{54}$H$_{72}$ cluster modeling a graphane fragment. Large and small balls represent
carbon and hydrogen atoms, respectively.

\newpage
\vskip 20mm
\includegraphics[width=\hsize,height=15cm]{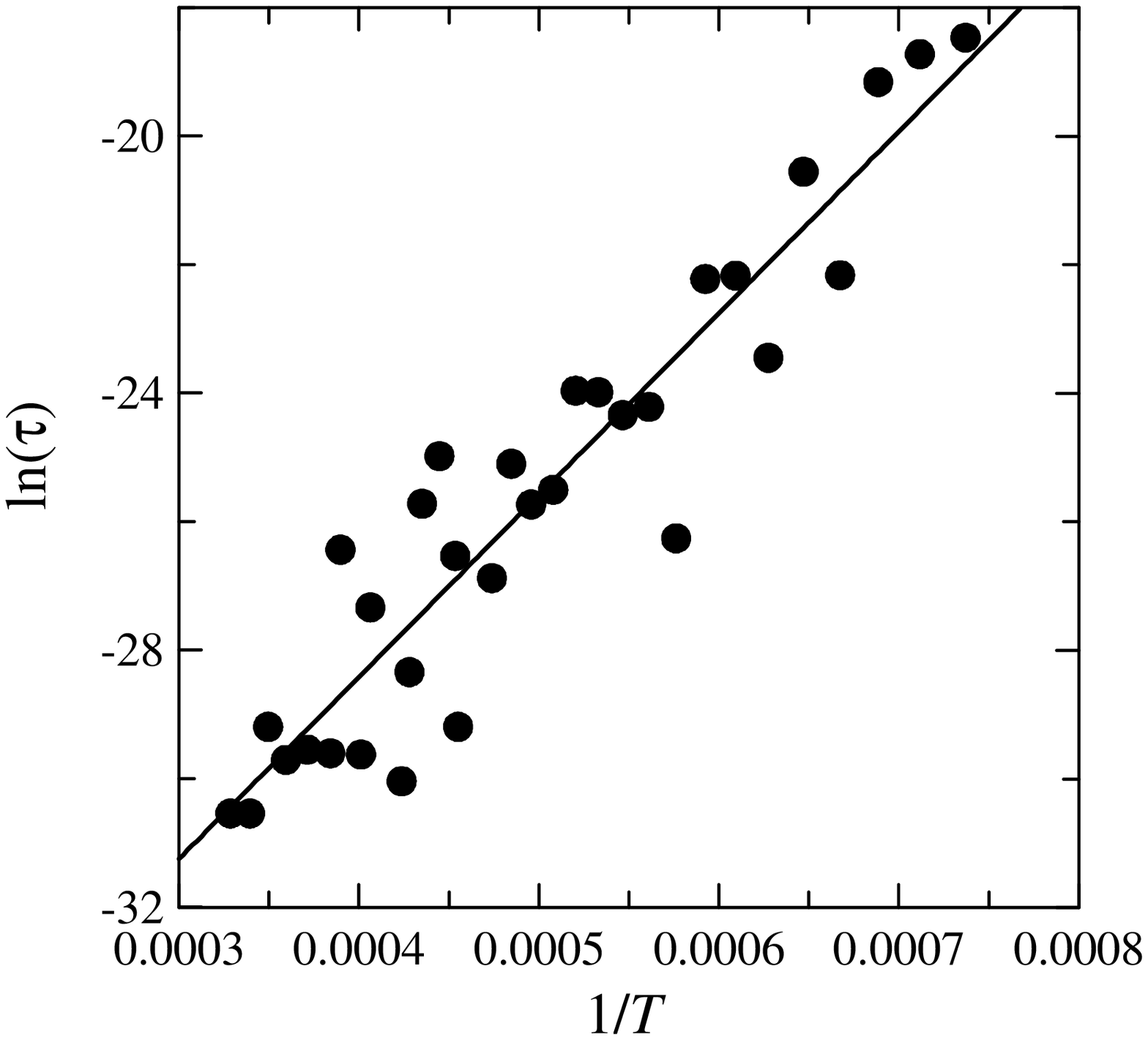}
\vskip 40mm
Fig. 2. Plot of the logarithm of desorption time $\tau$ [s] of one H atom from the C$_{54}$H$_{72}$
cluster versus inverse temperature $T$ [K]. Points present the results of model calculations,
solid line shows the linear approximation by least squares.


\begin{thebibliography}{10}
\bibitem{1} A. K. Geim and K. S. Novoselov, Nature Mater. $\mathbf{6}$, 183 (2007).
\bibitem{2} Z. Chen, Y.-M. Lin, M. J. Rooks, and P. Avouris,
Physica E $\mathbf{40}$, 228 (2007).
\bibitem{3} J. O. Sofo, A. S. Chaudhari, and G. D. Barber, Phys. Rev. B
$\mathbf{75}$, 153401 (2007).
\bibitem{4} D. C. Elias, R. R. Nair, T. M. G. Mohiuddin et al., Science $\mathbf{323}$, 610 (2009).
\bibitem{5} I. V. Davydov, A. I. Podlivaev, and L. A. Openov, Fiz. Tverd. Tela (St. Petersburg)
$\mathbf{47}$, 751 (2005) [Phys. Solid State $\mathbf{47}$, 778 (2005)].
\bibitem{6} L. A. Openov and A. I. Podlivaev, Pisma Zh. Eksp. Teor. Fiz. $\mathbf{84}$, 73 (2006)
[JETP Lett. $\mathbf{84}$, 68 (2006)].
\bibitem{7} M. M. Maslov, A. I. Podlivaev, and L. A. Openov, Phys. Lett. A {\bf 373}, 1653 (2009).
\bibitem{8} M. M. Maslov, D. A. Lobanov, A. I. Podlivaev, and L. A. Openov, Fiz. Tverd. Tela
(St. Petersburg) {\bf 51}, 609 (2009) [Phys. Solid State {\bf 51}, 645 (2009)].
\bibitem{9} A. I. Podlivaev and L. A. Openov, Pisma Zh. Eksp. Teor. Fiz. {\bf 81}, 656 (2005)
[JETP Lett. {\bf 81}, 533 (2006)].
\bibitem{10} J. Li, T. Furuta, H. Goto et al., J. Chem. Phys. $\mathbf{119}$, 2376 (2003).
\end{thebibliography}
\end{document}